\documentclass[apjl]{emulateapj}

\usepackage{amsmath}
\usepackage{natbib}
\usepackage{color}
\usepackage{graphicx}
\usepackage{xspace}
\usepackage[normalem]{ulem}

\newcommand{\Msun}{{\rm M_\odot}}
\newcommand{\hinvMsun}{h^{-1} {\rm M_\odot}}
\newcommand{\Mvir}{{\rm M_{vir}}}    
\newcommand{\dd}{{\rm {d}}}
\newcommand{\lpt}{\xspace{\small 2LPT}\xspace}
\newcommand{\za} {\xspace{\small ZA}\xspace}
\newcommand{\imf}{{\small IMF}\xspace}
\newcommand{\IGM}{{\small IGM}\xspace}
\newcommand{\rms}{{\rm rms}\xspace}

\newcommand{\moray}{{\sc Enzo+Moray}\xspace}
\newcommand{\popthree}{{Pop {\small III}}\xspace}
\newcommand{\BH}{\xspace{\small BH}\xspace}
\newcommand{\BHs} {\xspace{{\small BH}s}\xspace}
\newcommand{\DMH} {\xspace{{\small DMH}}\xspace}
\newcommand{\DMHs}{\xspace{{\small DMH}s}\xspace}

\definecolor{orange}{rgb}{1,0.5,0}

\let\oldmarginpar\marginpar
\renewcommand\marginpar[1]{\-\oldmarginpar[\raggedleft\footnotesize #1]%
{\raggedright\footnotesize #1}}

\shorttitle{2LPT and the First Black Holes}
\shortauthors{Holley-Bockelmann, Wise \& Sinha}

\begin{document}

\title{Kickstarting Reionization with the First Black Holes: the effects of second-order perturbation theory in pre-reionization volumes.}
\author{Kelly Holley-Bockelmann\altaffilmark{1,3}, John H. Wise\altaffilmark{2}, Manodeep Sinha\altaffilmark{1}}

\altaffiltext{1}{Department of Physics and Astronomy, Vanderbilt
  University, Nashville, TN, 37235; k.holley@vanderbilt.edu,
  manodeep.sinha@vanderbilt.edu}
\altaffiltext{2}{Center for Relativistic Astrophysics, Georgia
  Institute of Technology, Atlanta, GA, 30332; jwise@physics.gatech.edu}

\altaffiltext{3}{Adjunct at the Department of Physics, Fisk University, Nashville, TN, 37208}

\begin{abstract}
We explore structure formation in the dark ages ($z\sim 30-6$) using two well-known methods for 
initializing cosmological $N$-body simulations. Overall, both the Zel'dovich 
approximation (\za) and second order Lagrangian perturbation theory (\lpt) 
are known to produce accurate present-day dark matter halo mass functions. However, since the \lpt method drives
more rapid evolution of dense regions, it increases the occurrence of rare 
massive objects -- an effect that is most pronounced at high redshift. We find that \lpt 
produces more halos that could harbor Population~III stars
and their black hole remnants, and they produce them earlier. Although the differences between 
the \lpt and \za mass functions are nearly erased by $z=6$, this small boost to the number and mass of
black holes more than doubles
the reionized volume of the early Universe. We discuss the implications for reionization and massive black hole growth.
\end{abstract}


\maketitle

\section{Introduction}
The first generation of stars, so-called Population
III stars (\popthree), form from pristine gas within dark matter halos (\DMH) at very high redshifts
\citep{Couchman86, Tegmark97}.  Both one-zone calculations and
cosmological simulations suggest a top-heavy \popthree~\imf~with
an uncertain characteristic mass \citep[e.g.][]{Bromm:99, Omukai00,
  Abel:02firststar, Britton:2009}.  The uncertainty hinges on recent simulations 
  that showed the fragmentation and multiplicity may occur at protostellar densities~\citep{Turk:2009:firststars, Clark:2011:firststars,
  Greif:2012:firststars}, and on the idea that protostellar radiative
feedback may limit mass accretion to $\sim 40 \,\Msun$~\citep{Hosokawa11,
  Stacy12}.  Massive \popthree~stellar lifetimes are 
$\sim$ 2-20 Myr, after which they either enrich the intergalactic medium
(\IGM) with metals through supernovae~\citep{Heger:03sn}
or collapse directly into black holes (\BH) that could
grow into the supermassive ones at galactic centers~\citep{Madau:01Pop3, Volonteri:03smbh, Islam:03, taka09, Alvarez09, Jeon11, Micic:2011}.
While the \popthree~\imf~is still uncertain, it is likely that \DMHs do host
\popthree~stars, and a significant fraction produce seed \BHs.

To form the first generation of stars, primordial gas collapses
within the \DMH center via two separate
cooling channels -- molecular hydrogen and atomic hydrogen.
\citet{trenti:09} found the minimum \DMH mass to form
cold dense core through
molecular hydrogen cooling:
     
\begin{equation}
M_{\rm min} \simeq 6.44\times 10^6 \,\Msun J^{0.457}_{21} \Big({{1+z} \over {31}} \Big)^{-3.557},
\end{equation}
where $J_{21}$ relates to the Lyman-Werner flux, $F_\ast = 4 \pi J_{21} 10^{-21} \, {\rm erg} \, {\rm s^{-1}} \, {\rm cm ^{-2}}$.
It is this flux that is responsible for photo-dissociating molecular hydrogen. To calculate $F_\ast$, we adopt the comoving Lyman-Werner
photon density, $n_{\rm LW}$, from \citet{trenti:09}, which assumes stellar photon yields from \citet{schaerer:03}:

\begin{equation}
n_{\rm LW}(z)= 7\times10^{66} \, \times \, 10^{3.3-3.3(1+z)/11} \; \textrm{Mpc}^{-3},
\label{eqn:nlw}
\end{equation}

and

\begin{equation}
{J_{21}} = {1.6\times10^{-65} {\Bigg({{ n_{\rm LW} }\over {1 \, {\rm Mpc}^{-3}}} }\Bigg)} {\Bigg( {{1+z} \over {31}} \Bigg)^3}.
\end{equation}
    
\noindent
For $J_{21} \sim 5$ at
$z=10$, $M_{\rm min} \sim 5 \times 10^8\,\Msun$.~\footnote{Note that neutral HI in the IGM can reduce the 
effective LW flux from equation ~\ref{eqn:nlw} by an order of magnitude~\citep{Zoltan:2000}; this should not
affect our results since we are concerned with the {\em difference} between two otherwise identical volumes.}

On the other hand, all \DMHs with virial temperatures 
$\gtrsim 10^4$~K trigger efficient atomic hydrogen cooling, independent of 
the Lyman-Werner background. A virial temperature of $10^4$~K
translates into a \DMH mass of:
\begin{equation}
\Mvir = 7.75 \times 10^6 \,\Msun \left( \dfrac{1+z}{31} \right)^{-1.5}.
\end{equation}

Regardless of the cooling mechanism, the \DMHs that host 
the first generation of stars are by far
the most massive structures at this early epoch. For example,
a \DMH of $\sim 10^8 \, \Msun$ is a $\sim 5-\sigma$ peak at $z=20$.
These rare and massive peaks are the first collapsing structures in the Universe, and 
are the most likely to be plagued by numerical transients and initial value issues in any cosmological 
simulation. The purpose of this paper is to 
compare the effect on the \popthree era of two different techniques for initializing a 
cosmological $N$-body simulation: the Zel'dovich approximation (\za), and second 
order Lagrangian perturbation theory (\lpt) on the \popthree era. 
Below, we merely outline the cosmological initialization techniques; see \citet{Sco:1998} for a comprehensive
overview. 

The particle positions in an $N$-body simulation are generated from a density spectrum consistent
with the Cosmic Microwave Background (CMB). Since $N$-body simulations start at much lower redshifts ($z_{start}\sim50-250$),
the particle evolution up to the starting redshift is estimated using a displacement field.
In the Zel'dovich approximation, the displacement field assumes a linear evolution from recombination to
$z_{start}$ so the particle displacement is determined by the linear overdensity and linear 
growth factor $D_1$. In \lpt, the displacement field also includes a second order growth
factor $D_2\sim -3D^2_1/7$ and accounts for some of the non-linear density evolution.

Since \za is accurate only to first order, any higher order growing modes will be incorrect. In addition, 
in \za the non-linear decaying modes, called transients, damp away with $1/a$, while in \lpt, transients damp much
more quickly, as $1/a^2$.  This implies that the collapse time of true     
structures in an $N$-body simulation will be accurate after fewer
e-folding times when using \lpt~\citep[e.g.,][]{Sco:1998,crocce:06:2lpt, Jenkins:2010}.
Thus, for same starting redshift, simulations initialized with \lpt will capture
the formation of very high redshift halos more accurately compared to simulations initialized with \za.
In general, high-sigma peaks are supressed in \za, and the effect is larger at high redshift.
Of course, given enough expansion factors, the transients eventually damp out for all but the most massive
\DMHs. The standard lore is that a \za simulation must only start earlier than \lpt; 
indeed, by $z=1$ there is only a $\sim10\%$ difference in the \DMH number density  
above $10^{14}\,\Msun$~\citep{crocce:06:2lpt}. This small difference
in the {\em present day} \DMH mass function is perhaps why the Zel'dovich 
approximation has been adopted so widely.

\section{Simulations and Black Hole Modeling}\label{section:methods}
\subsection{Simulations}
In any cosmological simulation, the goal is to displace the particles
only slightly from their initial positions. The rms displacement,
$\Delta_\rms$ is:
\begin{equation}
\Delta_\rms^2  = \dfrac{4\pi}{3}\int_{k_f}^{k_{ny}} P(k,z_{start})\, dk ,
\end{equation}
\noindent where $k_f = 2\pi/L_{\rm box}$ is the fundamental mode, $L_{\rm box}$ is the simulation box-size,
$k_{ny}=N/2 \times k_f$ is the Nyquist frequency for an $N^3$ simulation, and $P(k,z_{start})$ is the power spectrum at the
starting redshift, $z_{start}$. If $\Delta_\rms$ is larger than the mean interparticle separation $\Delta_p = L_{\rm box}/N$,
then ``orbit crossing'' occurs and invalidates the accuracy of the initial conditions. Strictly speaking, imposing
$\Delta_\rms/\Delta_p=0.1$ would be ideal in that it makes orbit crossing a $\sim$ 10-sigma event.  The challenge in small volume, high resolution
simulations is that $\Delta_p$ is small and the starting redshift must be very
high to satisfy this $\Delta_\rms/\Delta_p$ constraint. For instance, to satisfy 
$\Delta_\rms/\Delta_p\sim 0.1 $, a 10 $h^{-1}$ Mpc, 512$^3$ simulation would need 
$z_{start}\approx799$. However, at very high redshifts the matter distribution is quite smooth, and the net force
may be dominated by numerical round-off errors~\citep{lukic:2007:halomf}  -- this will suppress small scale structure.
Therefore, we adopt  $\Delta_\rms/\Delta_p=0.25$, which sets $z_{start}=300$. 
We note that this criterion is rarely mentioned in the literature.  Several small volume, high resolution simulations
appear to have $\Delta_\rms/\Delta_p>1.0$, a clear sign that the initial conditions are not valid.
Since canonical \popthree-hosting halos are expected to appear at $z\sim30$, this $z_{start}$ allows 10 expansion factors to occur, which reduces transients from the initial conditions in \lpt by a factor of 100. 

We initialize six 512$^3$, 10 $h^{-1}$ Mpc, dark matter-only cosmological volumes
of a $\Lambda$CDM Universe and evolved them with Gadget-2~\citep{springel:01,springel:05}
from  $z_{start}=300$ to $z=6$.
The six simulations directly contrast the initialization 
technique, because each pair samples the volume identically from the CMB transfer 
function, and only displaces these identical initial positions to the 
same starting redshift using either \lpt or \za. Each volume is initialized 
using {\small WMAP}-5 parameters~\citep{Komatsu08}.

Given that we are concerned with the evolution of our volumes at very
high redshift, care was taken to integrate the positions and velocities with
high accuracy. We adopted many of the tree code parameters of the 
Coyote Universe~\citep{Heitmann:2010}. {\small PMGRID}, which defines
the Fourier grid is 1024, {\small ASMTH}, which defines the split between long and short-range 
forces, is $1.5$ times the mesh cell size, {\small RCUT}, 
which controls the maximum radius for short-range forces, 
is $6.0$ times the mesh cell size, the force accuracy is $0.002$, 
the integration accuracy is $0.00125$ and the softening is $1/40\Delta_p=0.5h^{-1}$ kpc.
The particle mass for all simulations is $5.3\times10^{5}\hinvMsun$.

\subsection{Black hole growth and accretion luminosity}
In post-processing, we identified halos with at least 20 particles
using {\small SUBFIND}~\citep{springel:01:subfind} and constructed a
halo mergertree~\citep{Sinha11}. We seeded $100 \,\hinvMsun$ \BHs in
halos with the mass criteria defined in the previous
section\footnote{Note that our particle mass does not allow us to
  resolve all \popthree-hosting halos at the highest redshifts}. Once
a \BH seed is sown, we do not allow another \BH to form within that
halo. The merger tree now allows us to track the growth and merger of
\BHs as well. We assumed a simple \BH growth prescription:
Eddington-limited accretion triggered by a major merger, where the
\BHs merge promptly after \DMHs merge, and there is no
  gravitational wave recoil~\citep[see][for more
  details]{micic:2006}.  While the \imf of \popthree stars and \BH
growth are still a matter of debate, the goal in this paper is simply
to examine the differences between \lpt and \za-seeded volumes.  A
different \imf or growth prescription will change absolute \BH number
densities and masses, and more (or less) aggressive \BH growth will
result in a difference in the ionized volumes. Hence, we chose the
simplest \popthree~\imf and \BH growth prescription to contrast the
two initialization techniques\footnote{We assume that the radiative
  feedback from the accreting \BH affects its growth by stopping the
  gas inflow after a Salpeter timescale}.

\begin{figure}[t]
\centering
\includegraphics[width=\columnwidth,clip=true]{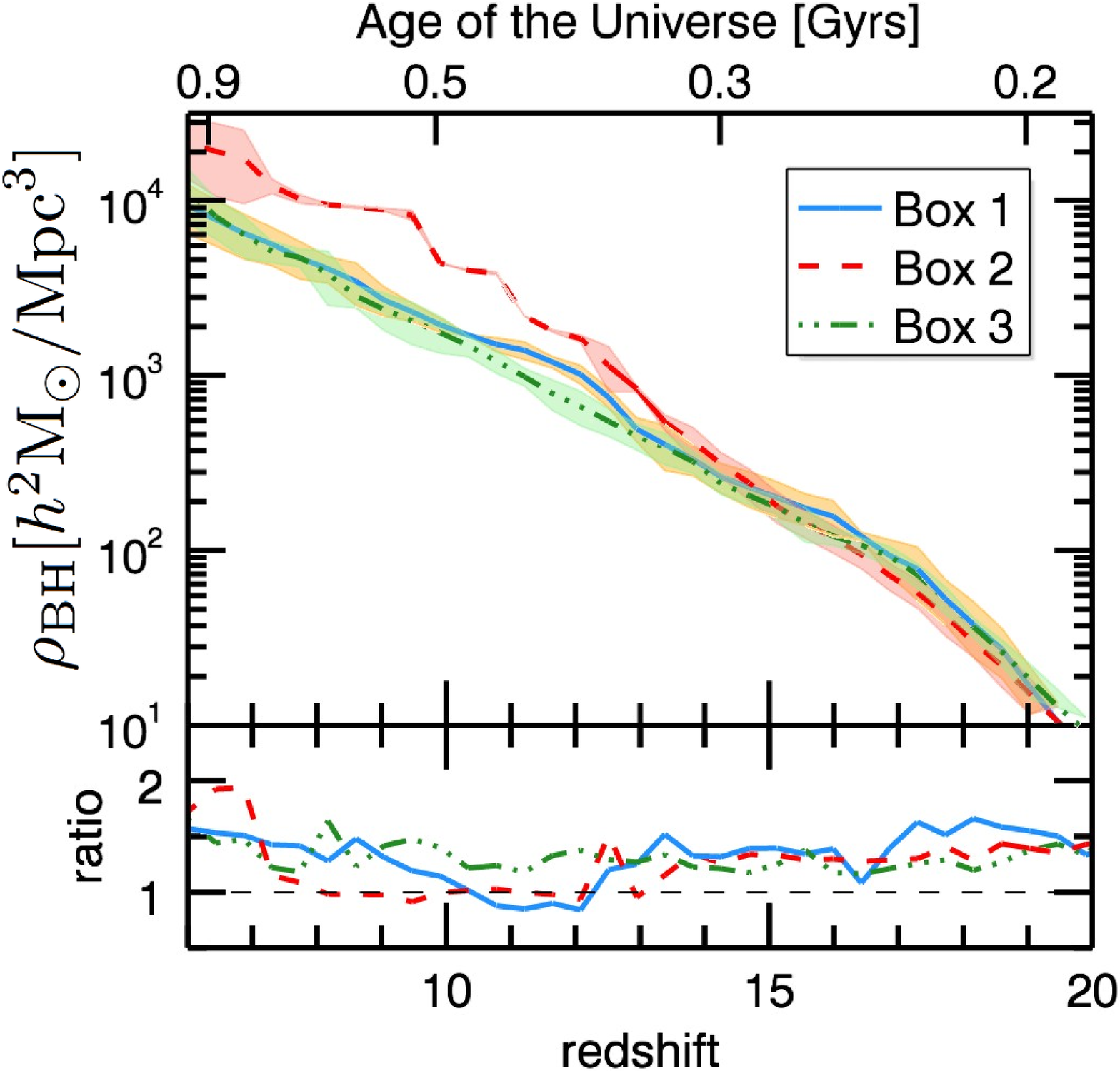}\hspace{0.5in}
\caption{\small Top: The co-moving \BH mass density as a function of redshift for  \lpt 
and \za for the three boxes. The lines represent the mean of the \lpt and \za \BH
mass density, while the shaded region shows the maximum and minimum of the \lpt and \za
quantities. Note the variation between the three boxes from cosmic variance. Bottom: the ratio of the \lpt and \za 
\BH masses as a function of redshift. \lpt \BHs
are more massive than the corresponding \za \BHs for the majority of cosmic time. }
\label{fig:blackholes}
\end{figure}

\begin{figure}
\centering
\includegraphics[width=\columnwidth,clip=true]{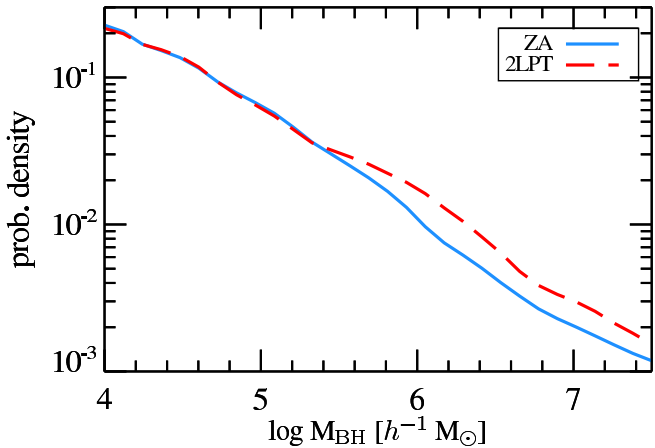}
\caption{\small  The probability density of the \BH mass function at $z=6$ computed by combining all 
    three boxes for the \lpt and \za initialized volumes.   
  The most massive black holes in the simulation volumes, $\log\, M_{\rm BH} \gtrsim 5.4$, are {\em consistently}
  more likely in the \lpt volumes by roughly 40\%. }
\label{fig:blackholes_mf}
\end{figure}

Once the \BHs are seeded and accreting, we estimate the ionizing
photon luminosity from the growing \BH, as well as its effect on the
neutral gas in the neighborhood through radiative transfer
calculations.  We assume that the \BH is fed at the Eddington rate by
major mergers, and this presupposes a rich supply of cold gas in these
halos -- an assumption that has not been observationally verified at
such a high redshift.  In fact, massive first stars can expel most of the gas in the host halo, starving the
remnant \BH~\citep{Kitayama04, Whalen04, Alvarez09} until a merger
with a gas-rich \DMH can restock it.  In particular, early simulations
with $\sim100\Msun$ \popthree~stars found that gas expulsion delayed
accretion for 10--50 Myr while the host \DMH restocked its gas
reservoir~\citep{Johnson07, Wise08_Gal, Wise12_Galaxy}.  Recent
simulations predict lower \popthree~stellar masses, and their
radiative feedback can still drive the majority of gas out of their
host halos.  However, most of the gas infall at high redshifts and in
rare peaks happen through cold streams -- these cold streams present a
small solid-angle to the stellar photon flux, which should make it
more difficult to photo-evaporate.  Thus, it may be plausible for
\popthree \DMHs, especially in more massive ones suppressed by LW radiation, 
to retain a reservoir of gas for future consumption.

Our simulation is collisionless, so we calculate the
impact of enhanced \BH growth on the neutral gas numerically in post-processing.
For an accreting \BH, the temperature profile of the accretion disk is:
\begin{align}
  T(r) &=  T_0  \left( \dfrac{r_{in}}{r}\right)^{3/4}  \left(1  -
  \sqrt{\dfrac{r_{in}}{r}} \,\right)^{1/4}, \\
  T_0  &= \left(\dfrac{3 G M_{\rm BH} \dot{M}_{gas}}{8 \pi \sigma
    r_{in}^3}\right)^{1/4}\notag,
\end{align}
where, $M_{\rm BH}$ is \BH mass, $\dot{M}_{gas}$ is gas accretion
rate, $\sigma$ is Stefan-Boltzmann's constant, $r_{in}$ is the
innermost accretion disk radius. The accretion disk luminosity
at a given frequency, $\nu$, can then be computed:
\begin{equation}
  \label{eqn:Lnu}
  L_\nu  = 4 \pi^2 \int_{r_{in}}^\infty \dfrac{2 h \nu^3}{c^2}
  \left[\exp\left(\dfrac{h\nu}{kT(r)}\right) -1 \right]^{-1} r \dd r
\end{equation}
where $L_\nu$ is the emitted luminosity per frequency band, $c$
is the speed of light, $k$ is Boltzmann's constant, and $h$ is
Planck's constant. The accretion efficiency, $\eta$ converts the accreted mass $(M_{acc})$
to energy: $M_{acc}\,c^2 = 1/2\, r_g/r_{in}$, where $r_g =GM_{\rm BH}/c^2$. 
For a Schwarzschild \BH $r_{in}$ is $\approx 6 r_g$ with
$\eta \sim 8\%$. In practice, we used $\eta=10\%$ and adjusted $r_{in}$ accordingly.

To quantify the effect of these \BHs on reionization, we conduct
radiative transfer calculations with \moray~\citep{wise:2011:moray}.
After every snapshot, we accumulate a fraction $\Omega_b/\Omega_M$ of
the mass into a fixed 512$^3$ grid, assuming all gas is hydrogen.  At
these scales and resolution, the gas
should follow the DM.  The density distribution is fixed during the
radiative transfer calculation.  We treat each accreting \BH as a
point source, with luminosities described by Equation (\ref{eqn:Lnu})
divided into bins of $13.6-40, 40-100, 100-10^3, 10^3-10^4$ eV.
Photon packages have the average photon energy in each bin, $\sim$ 28,
70, 400, and 1200 eV, and they obey periodic boundary conditions.  We
use a nonequilibrium chemistry model \citep{Anninos97} with hydrogen
only.  We consider Compton cooling and free electron heating by
the CMB, radiative losses from atomic cooling in the optically thin
limit, and secondary ionizations from high-energy radiation, using the
fitting formulae from \citet{Shull85}.

\section{Results}\label{section:results}
By $z=6$, we find that the \lpt-initialized volume produces 25%
more \popthree stars than the identical \za volume.  This is synonymous
with saying that high mass \DMHs collapsed earlier and more often. Indeed, at $z=24$,
there are 3 times the number of \DMHs above $10^7\Msun$, roughly a 5-$\sigma$ peak, and by $z=20$,
the difference in the halo mass function above $\sim 10^7\Msun$
($\sim4-\sigma$) is $\sim40\%$. Note that at $z=6$, the
difference in the number of {\it newly formed} \popthree star hosting
halos has virtually vanished -- this suggests that the simulation has
undergone enough e-folding times to damp away transients.

\begin{figure}[t]
\centering
\includegraphics[width=\columnwidth,clip=true]{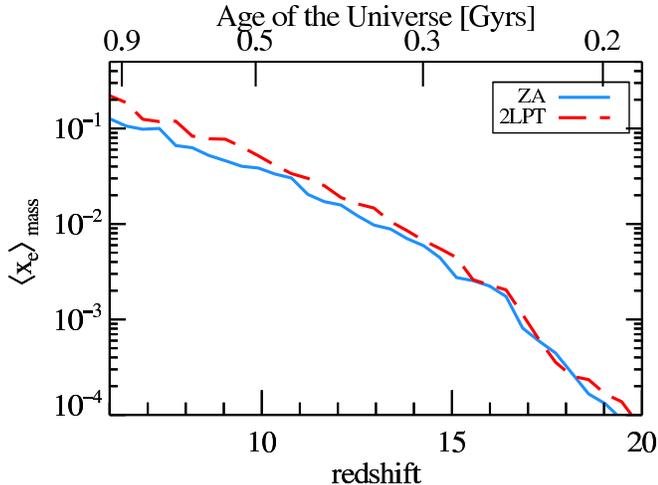}
\caption{\small Evolution of the  mass-weighted ionization fraction of the \za (solid) and
  \lpt (dashed) simulations.  The \lpt-initialized simulation generates
  20\% more \popthree-hosting halos, which enhances \BH 
  merger rates and thus their total luminosity and contribution to
  reionization.}
\label{fig:compare}
\end{figure}

\begin{figure}[t]
\centering
\includegraphics[width=\columnwidth,clip=true]{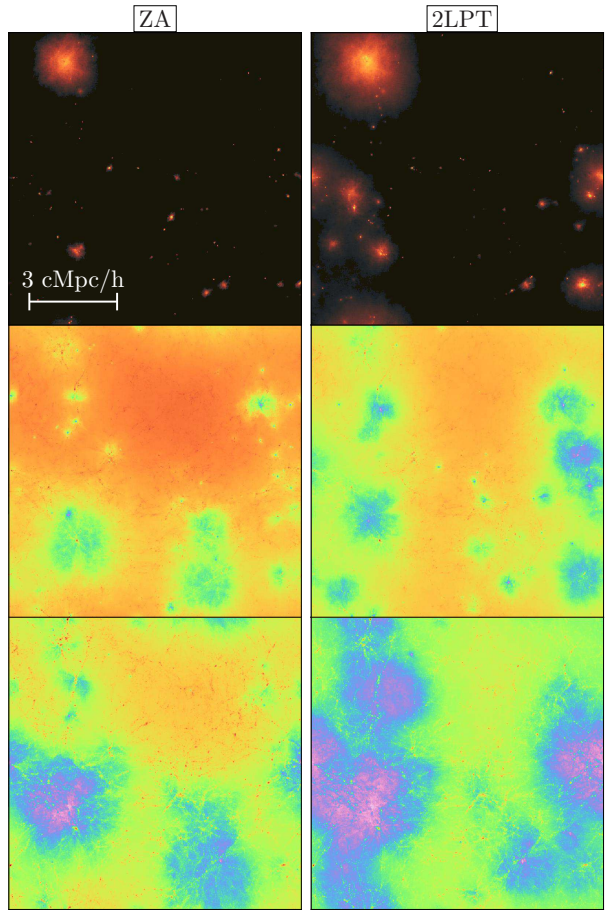}
\caption{\small The response of neutral hydrogen in the
  Universe to the radiation produced by seed \BHs in \za (left) and
  \lpt (right) initialized runs at three times in one realization. Both
  simulations used the same \BH seeding and growth mechanism, and sampled
  the same initial phases.  We used ray-tracing to propagate
  the BH radiation throughout the box, calculating the electron
  temperature of the gas and ionization state.}
\label{fig:reion}
\end{figure}

\subsection{The Changes in the Seed Black Hole Population}
An overall 25\% increase in the number of \popthree-hosting halos may 
not sound huge, but the consequences of this early and 
generous sowing of \popthree stars on proto-supermassive \BH growth is more profound. 
When we grow seed BHs with the major-merger driven prescription 
outlined above, the minor difference in the halo mass function magnifies. 
Figure~\ref{fig:blackholes} shows the \BH mass density, $\rho_{\rm BH}$, as a function of redshift. We 
calculate a factor of $\sim$ 2  difference in $\rho_{\rm BH}$ by $z=6$. If we allow a very vigorous, and perhaps unrealistic,  \BH growth scheme 
that fuels the \BH continuously at the Eddington rate, then $\rho_{\rm BH, \lpt}$ would be more than 10 times
 larger than the \za approximation. 
Future space-based gravitational wave observatories may detect these \BH inspirals; the gravitational wave signal in this 
pre-reionization era could feature more numerous and louder sources than predicted, depending on the detector configuration.

\subsection{Implications for the Reionization Epoch}\label{subsection:reionization}
Perhaps the most interesting consequence of the \lpt \DMH mass
function is the effect of the more numerous and more massive \BHs on
reionization. Figure~\ref{fig:blackholes_mf} shows the probability density of the \BH mass function at $z=6$ over all \lpt and \za initialized volumes. We find that massive black holes ($\log\,M_{\rm BH}\gtrsim5.4$) are $\sim1.4$
times more common in \lpt volumes. 

Although our simulation is purely collisionless, we can
estimate the contribution from the first \BHs on reionization. By
$z=6$, we find that \lpt \BHs emit 75\% more ionizing photons per
hydrogen atom than \za \BHs. This alone implies that the ionization history will be dramatically different. 
We next present the results from three-dimensional 
radiative transfer calculations on these six volumes.

Figure \ref{fig:compare} shows the evolution of the total \BH
luminosity and mass-weighted electron fraction in the \za and \lpt simulations in one realization.  At $z>10$, the total
luminosity fluctuates but is roughly equivalent.
Afterwards, the total luminosity in the \lpt simulation is consistently
$\sim3$ times greater than \za.  This results in
a larger ionized fraction at $z<9$, and yields a total
ionized fraction of 0.16 and 0.25 for \za and \lpt at $z=6$, respectively.  However, this enhanced contribution to reionization
only translates to a change in the Thomson optical depth of $\Delta
\tau_e=1.7 \times 10^{-3}$, if the universe is completely ionized at $z<6$.  In the other two realizations, the ionized fraction increases by 0.12 and 0.09 at $z=6$.

The density-weighted projections of electron fraction in Figure
\ref{fig:reion} depicts the spatial differences between \za and
\lpt at three redshifts.  The \ion{H}{2} regions around
the BHs are generally larger and more abundant in the \lpt case. Overall, the \lpt 
ionized volume is $\sim2.5$ times larger, where we define ionized as $x_e>0.5$.
Furthermore, this additional luminosity globally warms the \IGM outside 
these \ion{H}{2} to 17,000 K in underdense
regions ($\delta<1$), compared to 14,000 K in \za. This IGM global warming by 
pre-reionization \BHs may be important in regulating the formation and growth of
later, lower mass \BHs~\citep{takainprep}.

\section{Discussion}\label{section:discussion}
We found that pre-reionization simulations are particularly sensitive to
the phase-space initialization method, given
that the astrophysically interesting \DMHs -- ones that host the first
stars, \BHs, and protogalaxies -- are all expected to collapse at very
high redshifts. This means that the transient errors in the initial
positions and velocities have only a few e-folding times to damp away
before we need to make reliable measurements of the collapsed \DMHs. We found that the Zel'dovich approximation
underestimated the number and mass of the high mass \DMHs during the
dark ages, and this $\sim25\%$ difference in the extreme end of the
halo mass function magnified, increasing the \BH mass density $\sim 1.4$ 
times for $\log \, M_{\rm BH} \gtrsim 5.4$ at $z=6$ (see Fig.~\ref{fig:blackholes_mf}). 
This difference depends both on the fact that seed BHs form earlier, which allows them to
grow for a longer time through mergers and accretion, and also that they are sown more often.

One of the largest effects of this more accurate \lpt technique is
apparent in the reionization history of the early Universe. The volume of the Universe 
reionized at $z=6$ more than doubles, {\em only} due to the initialization
method.  We caution that our main goal was to study differences in
our two testbed simulations using simple prescriptions for \BH
feedback and mass accretion. We neglect cold gas inflows, mechanical
feedback, and indeed all gas and reionization physics is
treated in post-processing. We are also neglecting a potentially critical
effect of dark matter streaming motions on baryons at $z=1000$, which will also delay galaxy formation
compared to a standard approach~\citep{tsel:2010}. While the actual reionized volume of
universe is not robust here, the critical point is
that both \za and Press-Schechter approaches underestimate the
volume merely because they both assume a linear evolution of DM overdensities.

We focused on \popthree-star hosting halos here, simply because the
predicted \DMH mass thresholds are easily defined. 
However, a \za-initialized simulation  will always underestimate
the very high end of the \DMH mass function during the 
pre-reionization era, and this will have an effect on any prediction that
relies on accurate high mass \DMH number counts -- an obvious example 
is seed \BH formation from direct collapse~\citep{bromm:2003:ffs,lodato:2006:directsmbh,begelman:2006:directsmbh}.

Note, too, that our simulation is $\sim10^{6}$ times too small to model billion solar mass \BHs, whose number 
density is 1 per Gpc$^3$~\citep{Jiang:2009:smbh}. Although {\small WMAP}-5 can constrain the number of $\gtrsim 10^{13}\Msun$ halos, our
volume is too small to sample these halos. We do predict that these even rarer peaks will
be more numerous and will be, on average, more massive in the pre-reionization
epoch than predicted by N-body simulations initialized with \za. In principle, observations will be able to constrain
the true halo mass function in the dark ages. While the relatively
small difference in $\tau_e$ we observed may not be distinguishable with {\small PLANCK} and Herschel~\citep{Zahn11}, the
global IGM temperature difference may be detectable~\citep{Theuns02}, and SKA could easily probe this difference in
the 21cm power spectrum~\citep{Barkana:2010}.

\acknowledgments This work was conducted at the Advanced Computing Center 
for Research and Education at Vanderbilt University, Nashville, TN.  We also acknowledge 
support from NSF CAREER award AST-0847696. We would like to thank the referee for helpful comments. 


\newcommand{\noop}[1]{}

\end{document}